\documentclass{ws-p8-50x6-00}

\newcommand{\lsim}{\mathrel{\mathop{\kern 0pt \rlap
  {\raise.2ex\hbox{$<$}}}
  \lower.9ex\hbox{\kern-.190em $\sim$}}}
\newcommand{\gsim}{\mathrel{\mathop{\kern 0pt \rlap
  {\raise.2ex\hbox{$>$}}}
  \lower.9ex\hbox{\kern-.190em $\sim$}}}
\newcommand{\gagamma}{g_{a\gamma\gamma}}

\begin{document}

\title{Recent results from the canfranc dark matter search with
germanium detectors}












\author{
I.G.~Irastorza$^{\lowercase{a}}$\footnote{\lowercase{Attending
speaker: Igor.Irastorza@posta.unizar.es}},
A.~Morales$^{\lowercase{a}}$, C.E.~Aalseth$^{\lowercase{b}}$,
F.T.~Avignone~III$^{\lowercase{b}}$,
R.L.~Brodzinski$^{\lowercase{c}}$, S.~Cebri\'{a}n$^{\lowercase{a}}$,
E.~Garc\'{\i}a$^{\lowercase{a}}$, D.~Gonz\'{a}lez$^{\lowercase{a}}$,
W.K.~Hensley$^{\lowercase{c}}$, H.S.~Miley$^{\lowercase{c}}$,
J.~Morales$^{\lowercase{a}}$,
A.~Ortiz~de~Sol\'{o}rzano$^{\lowercase{a}}$,
J.~Puimed\'{o}n$^{\lowercase{a}}$, J.H.~Reeves$^{\lowercase{c}}$,
M.L.~Sarsa$^{\lowercase{a}}$, S.~Scopel$^{\lowercase{a}}$,
J.A.~Villar$^{\lowercase{a}}$}


\address{The COSME-IGEX joint Collaboration\\ ~\\
$^{a}$Laboratory of Nuclear and High Energy Physics, University of
Zaragoza, 50009 Zaragoza, Spain
\\
$^{b}$University of South Carolina, Columbia, South Carolina 29208
USA
\\
$^{c}$Pacific Northwest National Laboratory, Richland, Washington
99352 USA
}


\maketitle

\abstracts{Two germanium detectors are currently operating in the
Canfranc Underground Laboratory at 2450 m.w.e looking for WIMP
dark matter. One is a 2 kg $^{76}$Ge IGEX detectors (RG-2) which
has an energy threshold of 4 keV and a low-energy background rate
of about 0.3 c/keV/kg/day. The other is a small (234 g) natural
abundance Ge detector (COSME), of low energy threshold (2.5 keV)
and an energy resolution of 0.4 keV at 10 keV which is looking for
WIMPs and for solar axions. The analysis of 73 kg-days of data
taken by COSME in a search for solar axions via their photon
Primakoff conversion and Bragg scattering in the Ge crystal yields
a 95\% C.L. limit for the axion-photon coupling $g_{a \gamma
\gamma} < 2.8 \times 10^{-9}$  GeV$^{-1}$. These data, analyzed
for WIMP searches provide an exclusion plot for WIMP-nucleon
spin-independent interaction which improves previous plots in the
low mass region. On the other hand, the $\sigma(m)$ exclusion plot
derived from the 60 kg-days of data from the RG-2 IGEX detector
improves the exclusion limits derived from other ionization (non
thermal) germanium detector experiments in the region of WIMP
masses from 30 to 100 GeV recently singled out by the reported
DAMA annual modulation effect.}

\section{Introduction}

Substantial evidence exists suggesting most matter in the universe
is dark, and there are compelling reasons to believe it consists
mainly of non-baryonic particles. Among these candidates, Weakly
Interacting Massive Particles (WIMPs) and Axions are among the
front runners. The lightest stable particles of supersymmetric
theories, like the neutralino, describe a particular class of
WIMPs\cite{Gri}.

Direct detection of WIMPs rely on the measurement of their elastic
scattering off target nuclei in a suitable detector\cite{Mor99}.
Slow moving ($\sim300$ km/s) and heavy ($10 - 10^3$ GeV) galactic
halo WIMPs could make a Ge nucleus recoil with a few keV, at a
rate which depends on the type of WIMP and interaction. Ultra--low
background detectors with very low energy thresholds are needed
for such purpose. Germanium detectors have reached one of the
lowest background levels of any type of detector and  have a
reasonable ionization yield($\sim 0.25$). Thus, with sufficiently
low energy thresholds, they are attractive devices for WIMP direct
detection. In addition, solar axions can also been searched
\cite{Paschos:1994yf,Creswick:1998pg} with these crystal detectors
by looking for the Primakov conversion into photons and Bragg
diffraction.

Two experiments with germanium detectors are currently running at
the Canfranc Underground Laboratory: IGEX-DM and COSME-2. Data
obtained with both experiments are presented in this work, as well
as the WIMP exclusions obtained from the analysis of both sets of
data. COSME data has also been analyzed looking for the solar
axion signal mentioned above.

\section{COSME-2 and IGEX-DM experiments}

The IGEX experiment, optimized for detecting $^{76}$Ge double-beta
decay, has been described in detail in refs.\cite{Aal,Gon99}. The
IGEX detectors are now also being used in the search for WIMPs
interacting coherently with germanium nuclei. The COSME detector,
which has already been used in the past for dark matter
searhces\cite{Gar92}, is now operating in the same shield that
IGEX at a deeper location in Canfranc. Details about the features
of the detectors as well as the techniques used in their
construction can be found in the literature
\cite{Mor99,Gar92,Gon99,IGEX-DM}.

The IGEX detector used for dark matter searches, designated RG-II,
has an active mass of $\sim2.0$~kg. The full-width at half-maximum
(FWHM) energy resolution of RG-II was 0.8~keV at the 75-keV Pb
X-ray. The COSME detector has an active mass of 234~g and its FWHM
energy resolution is 0.43~keV at the 10.37~keV gallium X-ray.
Energy calibration and resolution measurements were made every
7--10 days using the lines of $^{22}$Na extrapolating to the low
energy region using the X-ray lines of Pb.


The detectors shielding is as follows, from inside to outside. The
innermost shield consists of 2.5 tons of 2000-year-old
archaeological lead ($^{210}$Pb~$<$~10~mBq/kg) forming a 60-cm
cube. The detectors fit into precision-machined holes in this
central core, which minimizes the empty space around the detectors
available to radon. Nitrogen gas, at a rate of 140~l/hour,
evaporating from liquid nitrogen, is forced into the detector
chambers to create a positive pressure and further minimize radon
intrusion. The archaeological lead block hosting the detectors is,
at its turn, surrounded by bricks ($\sim 10$~tons) of low activity
lead ($^{210}$Pb~$<$~30~Bq/kg) forming a total cube of 1 m side. A
minimum of 15~cm of archaeological lead separates the detectors
from the outer lead shield. A 2-mm-thick cadmium sheet surrounds
the main lead shield, and two layers of plastic seal this central
assembly against radon intrusion. A cosmic muon veto consisting of
plastic scintillators covers the top and sides of the central
core, except where the detector Dewars are located. An external
polyethylene neutron moderator 20~cm thick (1.5 tons) completes
the shield. The entire shield is supported by an iron structure
resting on noise-isolation blocks. The experiment is located in a
room isolated from the rest of the laboratory and has an
overburden of 2450 m.w.e., which reduces the measured muon flux to
$2 \times 10^{-7} \rm cm^{-2} \rm s^{-1}$.

\section{WIMP exclusion plot}

\begin{figure}[t]
\centerline{
\epsfxsize=8cm 
\epsfbox{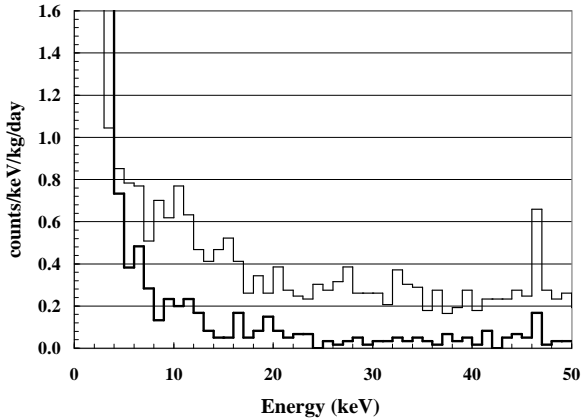} 
}

\caption{Low-energy spectrum of the 60 kg-d from IGEX RG-II
detector (thick line) and the 73 kg-d from COSME-2 detector (thin
line) \label{dm-ig-1}}
\end{figure}

\begin{figure}[t]
\centerline{
\epsfxsize=8cm 
\epsfbox{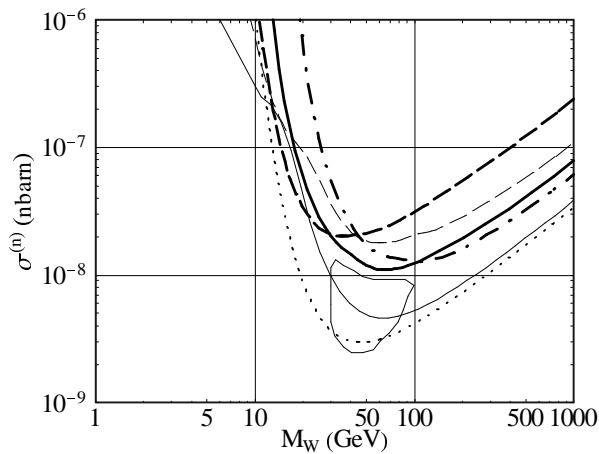} 
}

\caption{Exclusion plot for spin-independent interaction obtained
with IGEX-DM data (thick solid line) as well as COSME-2 data
(thick dashed line). Results obtained in other Germanium
experiments are also shown: Canfranc COSME-1
data\protect\cite{Gar92} (dot-dashed line) and the previous
Ge-combined bound (thin dashed line)
---including the last Heidelberg-Moscow data\protect\cite{Bau}.
The result of the DAMA NaI-0 experiment \protect\cite{Ber96} (thin
solid line) is also shown. The "triangle" area corresponds to the
(3$\sigma$) annual modulation effect reported by the DAMA
collaboration (including NaI-1,2,3,4
runnings)\protect\cite{Ber99}. The IGEX-DM projection (dotted
line) is shown for 1~kg-year of exposure with a background rate of
0.1~c/(keV-kg-day).\label{dm-ig-2}}
\end{figure}

The results presented in this talk correspond to 30 days (60 kg-d)
of data obtained with the IGEX RG-II detector as well as 311 days
(73 kg-d) obtained with COSME detector. Both spectra are plotted
in figure \ref{dm-ig-1}. The COSME spectrum shows a threshold of
2.5 keV and a low energy background (2.5 -- 10 keV) of 0.7
c/(keV-kg-day). On the other hand, the detector RG-II features an
energy threshold of 4 keV and the background rate recorded was
$\sim 0.37$ c/(keV-kg-day) between 4--10~keV, $\sim 0.12$
c/(keV-kg-day) between 10--20~keV, and $\sim 0.05$ c/(keV-kg-day)
between 20--40~keV.

The exclusion plots are derived from the recorded spectrum in
one-keV bins from threshold to 50~keV. The predicted signal in an
energy bin is required to be less than or equal to the (90\% C.L.)
upper limit of the (Poisson) recorded counts. The derivation of
the interaction rate signal supposes that the WIMPs form an
isotropic, isothermal, non-rotating halo of density $\rho =
0.3$~GeV/cm$^{3}$, have a Maxwellian velocity distribution with
$\rm v_{\rm rms}=270$~km/s (with an upper cut corresponding to an
escape velocity of 650~km/s), and have a relative Earth-halo
velocity of $\rm v_{\rm r}=230$~km/s. The cross sections are
normalized to the nucleon, assuming a dominant scalar interaction.
The Helm parameterization\cite{Eng91} is used for the scalar
nucleon form factor, and the ionization yield used is 0.25. The
exclusion plots derived from the IGEX-DM (RG-II) and COSME data
are shown in Fig.~\ref{dm-ig-2}. In particular, IGEX results
exclude WIMP-nucleon cross-sections above $1.3 \times 10^{-8}$ nb
for masses corresponding to the 50 GeV DAMA region\cite{Ber99}.
Also shown is the combined germanium contour, including the last
Heidelberg-Moscow data\cite{Bau} (recalculated from the original
energy spectra with the same set of hypotheses and parameters, but
with a suitable ionization yield\cite{Bau}), the DAMA experiment
contour plot derived from Pulse Shape Discriminated
spectra\cite{Ber96}, and the DAMA region corresponding to their
reported annual modulation effect\cite{Ber99}. The IGEX-DM
exclusion contour improves sizably on that of other germanium
ionisation experiments for masses corresponding to that of the
neutralino tentatively assigned to the DAMA modulation
effect\cite{Ber99} and results from using only unmanipulated data.

Based on present IGEX-DM performance and reduction of the
background to $\sim 0.1$~c/(keV-kg-day) between 4--10~keV,
Fig.~\ref{dm-ig-2} shows also the projection of IGEX for 1 kg-y of
exposure.

\section{Limit on axion-photon coupling}

Crystal detectors provide a simple mechanism for solar axion
detection \cite{Paschos:1994yf,Creswick:1998pg}. Axions can pass
in the proximity of the atomic nuclei of the crystal where the
intense electric field can trigger their conversion into photons.
In the process the energy of the outgoing photon is equal to that
of the incoming axion. The axion production rate in the Sun
--through Primakoff conversion of the blackbody photons in the
solar plasma-- can be easily estimated
\cite{vanBibber:1989ge,Creswick:1998pg} within the standard solar
model, resulting in an axion flux of average energy of about 4 keV
that can produce detectable X-rays in a crystal detector.
Depending on the direction of the incoming axion flux with respect
to the planes of the crystal lattice, a coherent effect can be
produced when the inverse momentum transfer equals the distance
between crystal planes (Bragg condition), leading so to a strong
enhancement of the signal. A correlation of the expected rate with
the position of the Sun in the sky follows providing a distinctive
signature of the axion which can be used, at the least, to improve
the signal/backgroung ratio.

A suitable statistical analysis \cite{Cebrian:1999mu}  to look for
this signature in the COSME data has been applied, and the
negative result leads to a limit on the axion-photon coupling
$g_{a\gamma \gamma} \lsim 2.8 \times 10^{-9} $ GeV$^{-1}$
practically equal to the one obtained by the SOLAX \cite{SOLAX}
Collaboration which is the (mass independent but solar model
dependent) most stringent laboratory bound for the axion-photon
coupling obtained so far for axion masses above 0.26 eV. Below
this mass value a recent result\cite{Tokyo} from a magnet
telescope (the Tokyo helioscope) improves a factor 3-5 the
$\gagamma$ COSME and SOLAX bounds.

\section*{Acknowledgements}
The IGEX results presented here have been obtained in
collaborative research with I.V.~Kirpichnikov, S.B.~Osetrov,
A.A.~Klimenko, A.A.~Smolnikov, A.A.~Vasenko, S.I.~Vasiliev,
V.S.~Pogosov and A.G.~Tamanyan, which have not participated,
however, in the COSME WIMP and axion searches. We thank them for
allowing the use of IGEX data already published in the open
scientific literature \cite{IGEX-DM}.


\end{document}